\documentclass[preprint,nofootinbib]{revtex4}%
\usepackage{amssymb}
\usepackage{amsfonts}
\usepackage{amsmath}
\usepackage{graphicx}%
\providecommand{\U}[1]{\protect\rule{.1in}{.1in}}
\begin{document}
\title{Field theories with anisotropic scaling in 2D, solitons and the microscopic
entropy of asymptotically Lifshitz black holes}
\author{Hern\'{a}n A. Gonz\'{a}lez$^{1}$, David Tempo$^{2}$, Ricardo Troncoso$^{2}$}
\affiliation{$^{1}$Departamento de F\'{\i}sica, Pontificia Universidad Cat\'{o}lica de
Chile, Casilla 306, Santiago 22, Chile.}
\affiliation{$^{2}$Centro de Estudios Cient\'{\i}ficos (CECs), Casilla 1469, Valdivia, Chile.}
\preprint{CECS-PHY-11/07}

\begin{abstract}
Field theories with anisotropic scaling in $1+1$ dimensions are considered. It
is shown that the isomorphism between Lifshitz algebras with dynamical
exponents $z$ and $z^{-1}$ naturally leads to a duality between low and high
temperature regimes. Assuming the existence of gap in the spectrum, this
duality allows to obtain a precise formula for the asymptotic growth of the
number of states with a fixed energy which depends on $z$ and the energy of
the ground state, and reduces to the Cardy formula for $z=1$.

The holographic realization of the duality can be naturally inferred from the
fact that Euclidean Lifshitz spaces in three dimensions with dynamical
exponents and characteristic lengths given by $z$, $l$, and $z^{-1}$,
$z^{-1}l$, respectively, are diffeomorphic. The semiclassical entropy of black
holes with Lifshitz asymptotics can then be recovered from the generalization
of Cardy formula, where the ground state corresponds to a soliton. An explicit
example is provided by the existence of a purely gravitational soliton
solution for BHT massive gravity, which precisely has the required energy that
reproduces the entropy of the analytic asymptotically Lifshitz black hole with
$z=3$.

Remarkably, neither the asymptotic symmetries nor central charges were
explicitly used in order to obtain these results.

\end{abstract}
\maketitle

\section{Introduction}

Field theories with anisotropic scaling are of interest in a variety of
subjects, specially in the context of quantum criticality in condensed matter
physics (see e.g. \cite{J.A.Hertz, Subir}). These theories have recently
attracted the attention of the high energy physics community with the aim of
finding a holographic description of the strong coupling regime, along the
lines of the AdS/CFT correspondence \cite{Herzog, Hartnoll, Maldacena
gauge-gravity}. In this sense, the so-called Lifshitz spacetimes, first
introduced by Kachru, Liu and Mulligan in \cite{KLM}, were proposed as
gravitational duals for which the anisotropic scaling symmetry is manifestly
realized though its isometries\footnote{These spacetimes were previously
discussed in the context of braneworlds in \cite{Koroteev-Libanov}. We thank
Peter Koroteev for bringing this reference to our attention.}. As in the case
of asymptotically AdS spacetimes \cite{Witten thermal}, finite temperature
effects are holographically introduced through a black hole in the bulk that
asymptotically approaches to the Lifshitz spacetime. Since then, several
asymptotically Lifshitz black hole solutions have been found, including
analytic
\cite{ALBHs0,ALBHs1,ALBHs2,ALBHs3,ABGGH00,ALBHs4,ABGGH,ALBHs6,ALBHs7,ALBHs8,ALBHs9}
and numerical results \cite{NLBHs1,NLBHs2,NLBHs3,NLBHs4,NLBHs5,NLBHs6,NLBHs7}.

In two spacetime dimensions, field theories with anisotropic scaling have also
been considered along different contexts in e.g., Refs. \cite{KLM12,KLM13,
Cardy-anisotropic}, and very recently in \cite{Hofman-Strominger}.

One of the main results reported here is that the semiclassical entropy of
black holes with Lifshitz asymptotics can be obtained from the asymptotic
growth of the number of states of a field theory with Lifshitz scaling in two
dimensions, where the ground state in the bulk corresponds to a soliton.

The plan of the paper is as follows. In the next Section, field theories with
Lifshitz scaling in $1+1$ dimensions are considered, and it is shown that a
duality between high and low temperature regimes naturally arises as a
consequence of the fact that Lifshitz algebras with dynamical exponents $z$
and $z^{-1}$ are isomorphic. In Section \ref{Cardy reloaded z}, a precise
formula for the asymptotic growth of the number of states with a fixed energy
which depends on $z$ and the energy of the ground state is found, which
reduces to the Cardy formula for $z=1$. The holographic realization of these
results is carried out in Sec. \ref{Holography}, where it is shown that the
semiclassical entropy of black holes with Lifshitz asymptotics can then be
recovered from the corresponding generalization of Cardy formula, provided the
ground state is identified as a soliton. An explicit example is provided in
Section \ref{Example BHT}, where a purely gravitational soliton solution for
BHT massive gravity is found, which precisely has the required energy that
reproduces the entropy of the analytic asymptotically Lifshitz black hole with
$z=3$ found in \cite{ABGGH00}, while an exhaustive computation for this case
is performed in Sec. \ref{Regularizaed action}. Finally, Section
\ref{Summary and discussion} is devoted to the summary and the discussion.

\section{Field theories with anisotropic scaling in 1+1 dimensions at finite
temperature}

Let us consider a field theory in two spacetime dimensions with anisotropic
scaling of the form,%
\begin{equation}
t\rightarrow\lambda^{z}t,\ x\rightarrow\lambda x\ . \label{DilationZ}%
\end{equation}
This symmetry together with translations and shifts in time given by
$x\rightarrow x+x_{_{0}}$, and $t\rightarrow t+t_{_{0}}$, are spanned by the
following algebra%
\begin{equation}
\lbrack P,H]=0,\ \ \ [D,P]=P,\ \ \ [D,H]=zH\ , \label{algebraLifshitzZ}%
\end{equation}
where $D$ is the generator of (\ref{DilationZ}), and $P$, $H$ stand for the
momentum generator and the Hamiltonian, respectively. This is known as a
two-dimensional Lifshitz algebra with dynamical exponent given by $z$ (see
e.g. \cite{Hartnoll, Maloney}).

A key observation that is worth to be stressed is the following: Lifshitz
algebras of the form (\ref{algebraLifshitzZ}) with dynamical exponents $z$ and
$z^{-1}$ are isomorphic. This can be easily verified by performing the
following change of basis%
\begin{equation}
\bar{P}=H,\ \ \ \ \bar{H}=P,\ \ \ \ \bar{D}=z^{-1}D\ , \label{ChangeOfBasis}%
\end{equation}
so that the algebra (\ref{algebraLifshitzZ}) is mapped into
\begin{equation}
\lbrack\bar{P},\bar{H}]=0,\ \ \ [\bar{D},\bar{P}]=\bar{P},\ \ \ [\bar{D}%
,\bar{H}]=z^{-1}\bar{H}\ . \label{algebraLifshitzZmenos1}%
\end{equation}
Remarkably, this local isomorphism naturally induces the equivalence between
the partition function at low and high temperatures for the same theory on a
cylinder of radius $l$. This can be seen as follows. At finite temperature
$T=\beta^{-1}$ the partition function can be defined on a torus where the
Euclidean time ($\tau=it$) is periodic, such that $0\leq\tau<\beta$, and
$0\leq\phi<2\pi l$. Therefore, the change of basis (\ref{ChangeOfBasis}) swaps
the role of Euclidean time and the angle, so that their periods would be
$0\leq\bar{\phi}<\beta$, and $0\leq\bar{\tau}<2\pi l$. The period of the angle
is then restored by virtue of scaling generated by $\bar{D}$ in
(\ref{algebraLifshitzZmenos1}), given by%
\begin{equation}
\bar{\tau}\rightarrow\lambda^{\frac{1}{z}}\bar{\tau},\ \bar{\phi}%
\rightarrow\lambda\bar{\phi}\ , \label{Scaling1/z}%
\end{equation}
with $\lambda=2\pi l\beta^{-1}$. Hence the period of $\bar{\tau}$ become
related with the period of $\tau$ according to
\begin{equation}
\bar{\beta}=\left(  2\pi l\right)  ^{1+\frac{1}{z}}\beta^{-\frac{1}{z}}\ ,
\label{Beta_Bar}%
\end{equation}
which allows to establish the relationship between the partition function at
low and high temperatures as $Z[\bar{\beta}]=Z\left[  \beta\right]  $. In sum,
the partition function can be assumed to be invariant under
\begin{equation}
Z\left[  \beta\right]  =Z\left[  \left(  2\pi l\right)  ^{1+\frac{1}{z}}%
\beta^{-\frac{1}{z}}\right]  \ , \label{Z-Duality}%
\end{equation}
which for $z=1$ reduces to the well known $S$-modular invariance for chiral
movers in conformal field theory \cite{Cardy}.

\subsection{Asymptotic growth of the number of states}

\label{Cardy reloaded z}

Assuming the existence of gap in the spectrum, the duality of low and high
temperatures regimes of the partition function (\ref{Z-Duality}) allows to
obtain a precise formula for the asymptotic growth of the number of states
with a fixed energy which depends on $z$ and the energy of the ground state,
which is assumed to be negative and given by $-\Delta_{0}$.

The gap guarantees that at low temperatures the partition function becomes
dominated by the contribution of the ground state, i.e., the partition
function can be approximated as $Z[\beta]\approx e^{\beta\Delta_{0}}$. By
virtue of Eq. (\ref{Z-Duality}) it follows that at high temperature the
partition function is given by $Z[\beta]\approx e^{(2{\pi l})^{1+\frac{1}{z}%
}\beta^{-\frac{1}{z}}\Delta_{0}}$, and therefore, the asymptotic growth of the
number of states at fixed energy $\Delta\gg\Delta_{0}$ can be obtained from%
\begin{align}
\rho(\Delta)  &  =\frac{1}{2\pi i}\int d\beta Z[\beta]e^{\beta\Delta
}\ ,\nonumber\\
&  \approx\frac{1}{2\pi i}\int d\beta e^{f(\beta,\Delta)}\ , \label{States}%
\end{align}
with $f(\beta,\Delta):=(2{\pi l})^{1+\frac{1}{z}}\beta^{-\frac{1}{z}}%
\Delta_{0}+\beta\Delta$. This expression can be evaluated in the saddle point
approximation which is characterized by the extremum of $f(\beta,\Delta)$.
This point corresponds to $\beta_{\ast}=2\pi l\left(  \frac{\Delta_{0}%
}{z\Delta}\right)  ^{\frac{z}{z+1}}$, and is such that\ $\left.
\partial_{\beta}f\right\vert _{\beta_{\ast}}=0$. The entropy is then given by
$S=\log\rho(\Delta)\approx f(\beta_{\ast},\Delta)$ which reads\footnote{We
have implicitly assumed that the ground state is non degenerate; otherwise the
asymptotic growth of the number of states is given by $\rho\left(
\Delta\right)  =\rho\left(  \Delta_{0}\right)  \exp\left[  2\pi l(z+1)\left[
\left(  \frac{\Delta_{0}}{z}\right)  ^{z}\Delta\right]  ^{\frac{1}{z+1}%
}\right]  $, where $\rho\left(  \Delta_{0}\right)  $ stands for the ground
state degeneracy.}%
\begin{equation}
S=2\pi l(z+1)\left[  \left(  \frac{\Delta_{0}}{z}\right)  ^{z}\Delta\right]
^{\frac{1}{z+1}}\ . \label{Cardyz}%
\end{equation}
Note that for $z=1$, the entropy reduces to $S=4\pi l\sqrt{\Delta_{0}\Delta}$,
so that Cardy formula is recovered\footnote{As explained in
\cite{TugaTroncoCristi}, in terms of the shifted Virasoro operator $\tilde
{L}_{0}=L_{0}-\frac{c}{24}$, Cardy formula can be expressed only through its
fixed and lowest eigenvalues, given by\ $\tilde{\Delta}$ and $-\tilde{\Delta
}_{0}$, respectively. Thus, the asymptotic growth of the number of states can
be obtained from the spectrum without making any explicit reference to the
central charges. The corresponding energies are then given by $\Delta
=\tilde{\Delta}l^{-1}$ and $-\Delta_{0}=-\tilde{\Delta}_{0}l^{-1}$,
respectively.}. Moreover, the entropy in (\ref{Cardyz}) fulfills the property
$S(\Delta_{0},\Delta,z)=S(\Delta,\Delta_{0},z^{-1})$, which is a consequence
of the isomorphism of the Lifshitz algebras (\ref{algebraLifshitzZ}) and
(\ref{algebraLifshitzZmenos1}).

In the Canonical ensemble, the first law, $d\Delta=TdS$, allows to express the
energy in terms of $\Delta_{0}$ and the temperature according to%
\begin{equation}
\Delta=\frac{1}{z}(2{\pi l})^{1+\frac{1}{z}}\Delta_{0}T^{1+\frac{1}{z}}\ ,
\label{Stefanboltzmann}%
\end{equation}
which reduces to the Stefan-Boltzmann law for $z=1$. Analogously, the entropy
can be written as
\begin{equation}
S=(2{\pi l})^{1+\frac{1}{z}}\left(  1+\frac{1}{z}\right)  \Delta_{0}%
T^{\frac{1}{z}}\ . \label{EntropyT}%
\end{equation}
In what follows it is shown that the thermodynamics of three-dimensional black
holes with Lifshitz asymptotics precisely fits the results found in this
section, where the ground state turns out to be a soliton.

\section{Anisotropic Holography, solitons and the microscopic entropy of
asymptotically Lifshitz black holes}

\label{Holography}

The holographic realization of the duality expressed by Eq. (\ref{Z-Duality})
can be naturally inferred from the fact that Euclidean Lifshitz spaces in
three dimensions with dynamical exponents and characteristic lengths given by
$z$, $l$, and $z^{-1}$, $z^{-1}l$, respectively, are diffeomorphic provided
their Euclidean time periods are related exactly as in Eq. (\ref{Beta_Bar}).
This can be seen as follows. The metric of Euclidean Lifshitz space in three
dimensions is given by \cite{KLM}%
\begin{equation}
ds^{2}=\frac{r^{2z}}{l^{2z}}d\tau^{2}+l^{2}\frac{dr^{2}}{r^{2}}+\frac{r^{2}%
}{l^{2}}d\phi^{2}\ , \label{EuclideanLifshitzMetric}%
\end{equation}
where $0\leq\tau<\beta$ and $0\leq\phi<2\pi l$. Performing a coordinate
transformation defined by%
\begin{equation}
\bar{\phi}=\frac{2\pi l}{\beta}\tau,\ \ \ \ \bar{\tau}=\left(  \frac{2\pi
l}{\beta}\right)  ^{\frac{1}{z}}\phi\ ,\ \ \ \ \bar{r}=\frac{\beta}{2\pi
z}\frac{r^{z}}{l^{z}}\ , \label{CoordinateLifsthitz}%
\end{equation}
makes the line element (\ref{EuclideanLifshitzMetric}) to acquire the form%
\begin{equation}
ds^{2}=\frac{z^{\frac{2}{z}}}{l^{\frac{2}{z}}}\bar{r}^{\frac{2}{z}}d\bar{\tau
}^{2}+\frac{l^{2}}{z^{2}}\frac{d\bar{r}^{2}}{\bar{r}^{2}}+\frac{z^{2}}{l^{2}%
}\bar{r}^{2}d\bar{\phi}^{2}\ , \label{Lifshitz1/zMetric}%
\end{equation}
where $0\leq\bar{\tau}<\bar{\beta}$ and $0\leq\bar{\phi}<2\pi l$, with
$\bar{\beta}=\left(  2\pi l\right)  ^{\frac{1+z}{z}}\beta^{-\frac{1}{z}}$ in
full agreement with Eq. (\ref{Beta_Bar}).

\bigskip

It is worth pointing out that, since this procedure is purely geometrical, the
result remains valid regardless the theory under consideration. Therefore,
analogously, an asymptotically Lifshitz Euclidean black hole with finite
temperature $\beta$ in three dimensions becomes diffeomorphic to a soliton
with temperature $\bar{\beta}$, with dynamical exponents and characteristic
lengths given by $z$, $l$, and $z^{-1}$, $z^{-1}l$, respectively. The
Lorentzian soliton is then regular everywhere and devoid of closed timelike
curves provided $\bar{t}=i\bar{\tau}$ is unwrapped. As a consequence the
soliton has a fixed mass, since the corresponding integration constant is
reabsorbed by a simple rescaling. It becomes then natural to regard the
soliton as the corresponding ground state.

Indeed, there are known examples for asymptotically AdS black holes and
solitons in three dimensions where this mechanism successfully reproduces the
microscopic entropy of the black hole by means of Cardy formula. This has been
explicitly shown for a different class of black holes with scalar hair and
scalar solitons in General Relativity \cite{TugaTroncoCristi, HMTZ-2+1}, as
well as for hairy black holes and solitons in vacuum for BHT massive gravity
\cite{PTT, GOTT, OTT}. In this sense, the soliton plays the role of AdS in
General Relativity \cite{Brown-Henneaux, Cardy, Strominger}, since Euclidean
AdS and the Euclidean BTZ black hole \cite{BTZ, BHTZ} are diffeomorphic
\cite{Carlip-Teitelboim, Maldacena-Strominger}.

Here we show how this mechanism can be extended in the case of asymptotically
Lifshitz black holes and solitons in three dimensions.

Let us consider a soliton with mass given by $M_{sol}=-M_{0}$, with $M_{0}>0$,
so that its Euclidean action reads%
\begin{equation}
I_{0}=\bar{\beta}M_{0}\ . \label{ISoliton_Z}%
\end{equation}
By virtue of the duality between low and high temperatures (\ref{Beta_Bar})
the value of the Euclidean action for the corresponding black hole is then
given by%

\begin{equation}
I=M_{0}(2{\pi l})^{1+\frac{1}{z}}\beta^{-\frac{1}{z}}\ . \label{IBH_Z}%
\end{equation}
Therefore, the black hole mass and its semiclassical entropy, given by
$M=-\partial_{\beta}I$, $S=\left(  1-\beta\partial_{\beta}\right)  I$,
respectively, turn out to be%
\begin{equation}
M=\frac{1}{z}(2{\pi l})^{1+\frac{1}{z}}M_{0}T^{1+\frac{1}{z}}\ , \label{Mbhz}%
\end{equation}
and%
\begin{equation}
S=(2{\pi l})^{1+\frac{1}{z}}\left(  1+\frac{1}{z}\right)  M_{0}T^{\frac{1}{z}%
}\ . \label{Sbhz}%
\end{equation}

Remarkably, the black hole mass (\ref{Mbhz}) and entropy (\ref{Sbhz})
precisely agree with the corresponding energy and entropy for the field theory
with dynamical exponent $z$ in $1+1$ dimensions, given by
(\ref{Stefanboltzmann}) and (\ref{EntropyT}), respectively, provided the black
hole mass is given by the energy of the field theory, and the energy of the
ground state is fixed by the soliton mass, i.e., for $\Delta_{0}=M_{0}$, and
$\Delta=M$.

It is then reassuring to verify that once formula (\ref{Mbhz}) is plugged into
(\ref{Sbhz}), one obtains%
\begin{equation}
S=2\pi l(z+1)\left[  \left(  \frac{M_{0}}{z}\right)  ^{z}M\right]  ^{\frac
{1}{z+1}}\ , \label{CardyzBH}%
\end{equation}
in full agreement with the asymptotic growth of the number of states in the
corresponding field theory, given by (\ref{Cardyz}).

For simplicity, we have considered here asymptotically Lifshitz spacetimes
whose only nonvanishing global charge is the mass. In presence of additional
global charges, one should work in the grand canonical ensemble.

An explicit example of these results is provided by the existence of a purely
gravitational soliton solution for BHT massive gravity, which precisely has
the energy that successfully reproduces the entropy of the analytic
asymptotically Lifshitz black hole with $z=3$ found in Ref. \cite{ABGGH00}.
This is discussed next.

\subsection{An explicit example for BHT massive gravity}

\label{Example BHT}

The new massive gravity theory, recently proposed by Bergshoeff, Hohm and
Townsend (\textbf{BHT}) is described by the following action \cite{BHT}%
\begin{equation}
I_{_{\text{\textrm{BHT}}}}=-\frac{1}{16\pi G}\int_{\mathcal{M}}d^{3}x\sqrt
{-g}\left[  R-2\lambda-\frac{1}{m^{2}}\left(  R_{\mu\nu}R^{\mu\nu}-\frac{3}%
{8}R^{2}\right)  \right]  \ . \label{ActionBHT}%
\end{equation}
Hereafter we will focus on a special case, given by%
\begin{equation}
\lambda=13m^{2}\ , \label{LifshitzPoint1}%
\end{equation}
for which the theory admits Lifshitz spacetimes as vacuum solutions with
dynamical exponents and characteristic lengths $z,l$ and $z^{-1},z^{-1}l$,
where $z=3$ and $l^{2}=-\frac{1}{2m^{2}}$. As pointed out in \cite{ABGGH00},
in this case the theory also admits a remarkably simple analytic
asymptotically Lifshitz black hole solution in vacuum with $z=3$, whose metric
reads%
\begin{equation}
ds^{2}=-\frac{r^{6}}{l^{6}}\left(  1-\frac{r_{+}^{2}}{r^{2}}\right)
dt^{2}+\frac{l^{2}}{r^{2}}\left(  1-\frac{r_{+}^{2}}{r^{2}}\right)
^{-1}dr^{2}+\frac{r^{2}}{l^{2}}d{\phi}^{2}\ . \label{LifshitzBH}%
\end{equation}
The coordinates range as $-\infty<t<+\infty$, $0<r<\infty$, $0\leq\phi<2\pi
l$, and the event horizon is located at $r=r_{+}$, surrounding the singularity
at the origin. Its Hawking temperature is then given by%
\begin{equation}
T=\beta^{-1}=\frac{1}{2\pi l}\left(  \frac{r_{+}}{l}\right)  ^{3}\ .
\label{T-BH}%
\end{equation}

\subsubsection{Gravitational Soliton}

The theory (\ref{ActionBHT}) at the special case (\ref{LifshitzPoint1}) also
turns out to admit a gravitational soliton in vacuum. The solution is given by%
\begin{equation}
ds^{2}=-\cosh^{2}\rho d\bar{t}^{2}+l^{2}d\rho^{2}+\cosh^{4}\rho\sinh^{2}%
\rho\ d\bar{\phi}^{2}\ , \label{LifshitzSoliton}%
\end{equation}
with $-\infty<\bar{t}<+\infty$, $0\leq\rho<\infty$, and $0\leq\bar{\phi}<2\pi
l$, and the scalar curvature reads%
\[
R=-\frac{18}{l^{2}}-\frac{8}{l^{2}}\tanh^{2}\rho\ .
\]
This spacetime is regular everywhere, geodesically complete and shares the
same causal structure with AdS spacetime. The soliton describes an
asymptotically Lifshitz spacetime with dynamical exponent $z=\frac{1}{3}$ and
characteristic length $\frac{l}{3}$, as it can be explicitly verified
performing a redefinition of the radial coordinate according to%
\begin{equation}
\bar{r}=\frac{l}{3}\cosh^{3}\rho\ , \label{r barra pho soliton}%
\end{equation}
so that for $\bar{r}\rightarrow\infty$, the metric (\ref{LifshitzSoliton})
approaches to that of a (Lorentzian) Lifshitz spacetime
(\ref{EuclideanLifshitzMetric}) with the corresponding dynamical exponent and
characteristic length.

It is worth highlighting that, unlike the Lifshitz spacetime
(\ref{EuclideanLifshitzMetric}), the soliton (\ref{LifshitzSoliton}) is devoid
of divergent tidal forces near the origin.

\bigskip

It is then simple to verify that the Euclidean continuation of the soliton
(\ref{LifshitzSoliton}) with $\bar{\tau}=i\bar{t}$, is diffeomorphic to the
Euclidean black hole (\ref{LifshitzBH}) with $\tau=it$, provided their
temperatures are related as in Eq. (\ref{Beta_Bar}). The corresponding
coordinate transformation is given by (\ref{CoordinateLifsthitz}), which in
order to fit the gauge choice of the radial coordinate as in Eq.
(\ref{LifshitzSoliton}), has to be followed by (\ref{r barra pho soliton}).

\bigskip

In sum, the theory (\ref{ActionBHT}) at the special case (\ref{LifshitzPoint1}%
) admits nontrivial asymptotically Lifshitz solutions in vacuum, given by the
black hole (\ref{LifshitzBH}) and the gravitational soliton
(\ref{LifshitzSoliton}). Note that since their corresponding dynamical
exponents and characteristic lengths are given by $3$, $l$, and $\frac{1}{3}$,
$\frac{l}{3}$, respectively, both metrics do not match at infinity, which
\textit{a priori }may appear as an obstacle to compare them in the same
footing. Indeed, the obstruction remains for asymptotically Lifshitz
spacetimes with generically different dynamical exponents and characteristic
lengths. Nonetheless, and remarkably, when they are related as $z$,$l$, and
$z^{-1}$, $z^{-1}l$, as in our case, the obstacle can be circumvented due to
the fact that their corresponding Euclidean continuations are diffeomorphic
provided the temperatures are related according to Eq. (\ref{Beta_Bar}), and
hence, any suitably regularized Euclidean action for the black hole is
necessarily finite for the gravitational soliton and vice versa. This is
explicitly confirmed in the next subsection.

\subsection{Regularized Euclidean action: Soliton as the ground state and
microscopic black hole entropy}

\label{Regularizaed action}

The BHT action (\ref{ActionBHT}) at the special point (\ref{LifshitzPoint1})
can be regularized following the quasilocal\ approach of Brown and York
\cite{Brown-York}, endowed with suitable counterterms along the lines of
\cite{Henningson-Skenderis, Balasubramanian-Kraus}. For the case under
consideration this task was carried out by Hohm and Tonni in \cite{Hohm-Tonni}%
, where it was shown that the regularized Brown-York stress energy tensor
gives the mass for the asymptotically Lifshitz black hole (\ref{LifshitzBH})
which agrees with the first law of thermodynamics. Here we show that the
regularized Euclidean action also reproduces the expected results, as in Eqs.
(\ref{ISoliton_Z}) and (\ref{IBH_Z}), for the gravitational soliton
(\ref{LifshitzSoliton}) and the black hole (\ref{LifshitzBH}), respectively.

In second order formalism the bulk action (\ref{ActionBHT}) can be expressed
as%
\begin{equation}
I_{_{\text{\textrm{BHT}}}}=-\frac{1}{16\pi G}\int_{\mathcal{M}}d^{3}x\sqrt
{-g}\left[  R-2\lambda-f^{\mu\nu}G_{\mu\nu}+\frac{1}{4}m^{2}\left(  f_{\mu\nu
}f^{\mu\nu}-f^{2}\right)  \right]  \ ,
\end{equation}
where $f^{\mu\nu}$ is an auxiliary field. The regularized action is given by
\cite{Hohm-Tonni}%
\begin{equation}
I_{\text{\textrm{reg}}}=I_{_{\text{\textrm{BHT}}}}+I_{_{\text{\textrm{GH}}}%
}+I_{\text{\textrm{ct}}}\ , \label{I-reg}%
\end{equation}
where the boundary terms read%
\begin{align}
I_{_{\text{\textrm{GH}}}}  &  =-\frac{1}{16\pi G}\int_{\partial\mathcal{M}%
}d^{2}x\sqrt{-\gamma}\left[  -2K-\hat{f^{ij}}K_{ij}+\hat{f}K\right]  \ ,\\
I_{\text{\textrm{ct}}}  &  =\frac{1}{32\pi Gl^{2}}\int_{\partial\mathcal{M}%
}d^{2}x\sqrt{-\gamma}\left[  15+\frac{1}{2}\hat{f}-\frac{1}{16}\hat{f}%
^{2}\right]  \ .
\end{align}
Here, $\hat{f^{ij}}$ is defined in terms of $f^{ij}$ and the shift $N^{j}$ of
a radial ADM decomposition of the bulk metric according to%
\begin{equation}
\hat{f^{ij}}=f^{ij}+2f^{r(i}N^{j)}+f^{rr}N^{i}N^{j}\ .
\end{equation}

\subsubsection{Gravitational Soliton}

Evaluating the Euclidean continuation of the gravitational soliton
(\ref{LifshitzSoliton}) on each term of the regularized action (\ref{I-reg}),
one obtains that the relevant terms are given by%
\begin{equation}
I_{_{\text{\textrm{BHT}}}}=\frac{\bar{\beta}}{2G}\left(  -\frac{e^{2\rho}}%
{2}+1\right)  \ ,\ I_{_{\text{\textrm{GH}}}}=\frac{\bar{\beta}}{2G}e^{2\rho
}\ ,\ I_{\text{\textrm{ct}}}=\frac{\bar{\beta}}{4G}\left(  -e^{2\rho
}+1\right)  \ ,
\end{equation}
so that the divergences exactly cancel out and the total Euclidean action
(\ref{I-reg}) for the soliton becomes finite and given by%
\begin{equation}
I_{0}=\frac{3}{4G}\bar{\beta}\ . \label{Isol}%
\end{equation}
Therefore, as expected, the soliton has a fixed negative mass is given by%
\begin{equation}
M_{sol}=-M_{0}=-\partial_{\bar{\beta}}I_{0}=-\frac{3}{4G}\ . \label{M-Soliton}%
\end{equation}

As explained in section \ref{Holography}, since the soliton
(\ref{LifshitzSoliton}) is diffeomorphic to the black hole (\ref{LifshitzBH}),
the Euclidean action of the latter is obtained from (\ref{Isol}), by virtue of
the relationship between their Euclidean time periods as in Eq.
(\ref{Beta_Bar}). Nevertheless, an explicit evaluation of the regularized
Euclidean action turns out to be a reassuring and healthy exercise which is
performed next.

\subsubsection{Asymptotically Lifshitz black hole}

Each term of the regularized action (\ref{I-reg}), once evaluated in the
Euclidean continuation of the black hole (\ref{LifshitzBH}), gives%
\begin{equation}
I_{_{\text{\textrm{BHT}}}}=\frac{\beta r_{+}^{2}}{Gl^{4}}\left(  -r^{2}%
+r_{+}^{2}\right)  \ ,\ I_{_{\text{\textrm{GH}}}}=\frac{\beta r_{+}^{2}%
}{Gl^{4}}\left(  2r^{2}-r_{+}^{2}\right)  \ ,\ I_{\text{\textrm{ct}}}%
=\frac{\beta r_{+}^{2}}{Gl^{4}}\left(  -r^{2}+\frac{3r_{+}^{2}}{4}\right)  \ ,
\end{equation}
and hence, divergences again cancel out so that the Euclidean action
(\ref{I-reg}) is finite and reads%
\begin{equation}
I=\frac{3\beta}{4G}\frac{r_{+}^{4}}{l^{4}}\ . \label{I-reg-bh}%
\end{equation}
Once expressed in terms of the temperature (\ref{T-BH}), the Euclidean action
(\ref{I-reg-bh}) is given by%
\begin{equation}
I=\frac{3}{4G}\left(  2\pi l\right)  ^{\frac{4}{3}}\beta^{\frac{-1}{3}}\ ,
\label{I-reg-bh-beta}%
\end{equation}
which reduces to $-\beta$ times the free energy in the semiclassical
approximation, i.e., $I=-\beta F$, with $F=M-TS$.$\ $Therefore, the first law
is recovered requiring the action to possess an extremum, so that the mass,
$M=-\partial_{\beta}I$, is given by%
\begin{equation}
M=\frac{1}{4G}\left(  2\pi l\right)  ^{\frac{4}{3}}\beta^{-\frac{4}{3}}%
=\frac{1}{4G}\left(  \frac{r_{+}}{l}\right)  ^{4}\ . \label{M-BH-Z3}%
\end{equation}
This is in agreement with the result found through the evaluation of the
Brown-York stress-energy tensor in \cite{Hohm-Tonni}. The mass has also been
computed by different methods in Refs.
\cite{LifshitzBH-Mass1,LifshitzBH-Mass2}. The semiclassical black hole entropy
is then given by $S=(1-\beta\partial_{\beta})I$, which reduces to%
\begin{equation}
S=\frac{(2\pi l)^{\frac{4}{3}}}{G}\beta^{-\frac{1}{3}}=\frac{2\pi r_{+}}{G}\ ,
\label{S-bh-Z3}%
\end{equation}
and agrees with the result found in \cite{ALBHs3,Hohm-Tonni} by means of
Wald's formula \cite{Wald}.

Note that by virtue of the precise value of the soliton mass in Eq.
(\ref{M-Soliton}), the Euclidean action, mass and black hole entropy obtained
in the previous section, given by the generic formulae (\ref{IBH_Z}),
(\ref{Mbhz}) and (\ref{Sbhz}), exactly reduce to the ones computed above,
given by (\ref{I-reg-bh}), (\ref{M-BH-Z3}) and (\ref{S-bh-Z3}), respectively.

The link with the microscopic counting of states of the field theory with
anisotropic scaling is then seen as follows. Once formula (\ref{M-BH-Z3}) is
plugged into (\ref{S-bh-Z3}) one obtains the black hole entropy as a function
of its mass, given by%
\begin{equation}
S=8\pi l\left[  \left(  \frac{1}{4G}\right)  ^{3}M\right]  ^{\frac{1}{4}},
\label{S-BH-Z3-M}%
\end{equation}
which precisely agrees with (\ref{CardyzBH}), and hence with and the
asymptotic growth of the number of states in Eq. (\ref{Cardyz}), provided the
black hole mass is identified with the energy of the dual field theory, and
the energy of the ground state is determined by the soliton mass, i.e., for
$\Delta_{0}=M_{0}=\frac{3}{4G}$ and $\Delta=M$.

\section{Summary and discussion}

\label{Summary and discussion}

Field theories with anisotropic scaling of the form (\ref{DilationZ}) in $1+1$
dimensions were analyzed. It was shown that the isomorphism between Lifshitz
algebras (\ref{algebraLifshitzZ}) and (\ref{algebraLifshitzZmenos1}) with
dynamical exponents $z$ and $z^{-1}$, respectively, naturally induces the
equivalence between the partition function at high and low temperatures for
the same theory on a cylinder of radius $l$, according to Eq. (\ref{Z-Duality}%
). In the case of $z=1$ it is simple to verify that the Lifshitz algebra
(\ref{algebraLifshitzZ}) is isomorphic to Poincar\'{e} in $1+1$, and it is
then amusing to verify that the well-known $S$-modular invariance for chiral
movers in conformal field theory can be recovered requiring much less
symmetries. The existence of a gap in the spectrum guarantees that at low
temperatures the partition function becomes dominated by the contribution of
the ground state. Hence, the duality allows to obtain a precise formula for
the asymptotic growth of the number of states with a fixed energy, given by
(\ref{Cardyz}), which depends on $z$ and the energy of the ground state, and
reduces to Cardy formula for $z=1$. In the canonical ensemble, the energy and
the entropy acquire the expected dependence on the temperature for a field
theory with the anisotropic scaling (\ref{DilationZ}) (see e.g. \cite{ALBHs0},
\cite{Kraus-D'Hoker}), and remarkably their precise form, which gives a
measure of the number of degrees of freedom, becomes determined by the
dynamical exponent $z$ and the ground state energy $-\Delta_{0}$ as in Eqs.
(\ref{Stefanboltzmann}) and (\ref{EntropyT}), respectively.

The duality expressed by Eq. (\ref{Z-Duality}) admits an interesting
holographic realization. This can be seen from the fact that Euclidean
Lifshitz spaces in three dimensions (\ref{EuclideanLifshitzMetric}) and
(\ref{Lifshitz1/zMetric}) with dynamical exponents and characteristic lengths
given by $z$, $l$, and $z^{-1}$, $z^{-1}l$, respectively, are related by the
coordinate transformation (\ref{CoordinateLifsthitz}), so that they are
diffeomorphic provided their Euclidean time periods are related exactly as in
Eq. (\ref{Beta_Bar}).

Remarkably, this procedure is purely geometrical, and hence the result remains
the same regardless the theory under consideration. As a consequence, the
semiclassical entropy of black holes with Lifshitz asymptotics can then be
recovered from the generalization of Cardy formula in (\ref{Cardyz}), where
the ground state corresponds to a soliton. This result stems from the fact
that an asymptotically Lifshitz Euclidean black hole in three dimensions and a
soliton, with dynamical exponents and characteristic lengths given by $z$,
$l$, and $z^{-1}$, $z^{-1}l$, respectively, become diffeomorphic provided
their Euclidean time periods are related as in Eq. (\ref{Beta_Bar}). In
particular, in the sense of Horava and Melby-Thompson
\cite{Horava+Melby-Thompson}, the anisotropic conformal boundary of the
Euclidean black hole and the soliton is then the same torus.

The Euclidean action of the black hole (\ref{IBH_Z}) is then related to the
one of the soliton (\ref{ISoliton_Z}) by virtue of the duality between high
and low temperatures (\ref{Beta_Bar}), and therefore, the black hole entropy
is shown to be given by (\ref{CardyzBH}). Remarkably, this formula precisely
agrees with the asymptotic growth of the number of states in the corresponding
field theory with dynamical exponent $z$ in $1+1$ dimensions, given by
(\ref{Cardyz}), provided the black hole mass is given by the energy of the
field theory, and the energy of the ground state is fixed by the soliton mass.

\bigskip

Remarkably, neither the asymptotic symmetries nor central charges were
explicitly used in order to obtain these results, and hence they should
naturally extend for higher dimensional black holes with an emergent
three-dimensional Lifshitz geometry near the horizon.

\bigskip

Since the soliton mass is negative, the specific heat of the black hole
\begin{equation}
C=\frac{\partial M}{\partial T}=\frac{1+z}{z^{2}}(2{\pi l})^{1+\frac{1}{z}%
}M_{0}T^{\frac{1}{z}}\ , \label{Specific heat}%
\end{equation}
is manifestly positive, and therefore the black hole may reach local thermal
equilibrium with a heat bath. Nonetheless, at the critical temperature
$T=\frac{1}{2\pi l}$, which corresponds to the self-dual point of the
transformation (\ref{Beta_Bar}), i.e., $\bar{\beta}=\beta$, according to Eqs.
(\ref{ISoliton_Z}) and (\ref{IBH_Z}), the soliton and the asymptotically
Lifshitz black hole possess the same free energy. This means that at high
temperatures the partition function turns out to be dominated by the black
hole, while for low temperatures it becomes dominated by thermal radiation on
the soliton. Note that at fixed temperature, apart from the black hole and the
gravitational soliton, the Lifshitz spacetime also appears as an additional
possible decay channel. However, since its free energy vanishes, the Lifshitz
spacetime cannot dominate the partition function and it is then always
unstable against thermal decay.

It is worth pointing out here that at an arbitrarily small temperature, since
the horizon radius of the black hole becomes arbitrarily small, an external
observer that is nearby the horizon may feel arbitrarily large tidal forces,
as it occurs for a Lifshitz spacetime around the origin \cite{Horowitz-Ross,
KLM, Copsey-Mann}. However, as explained above, since in this case the soliton
is the preferred configuration, this mechanism can be regarded as a sort of
extension of cosmic censorship that resolves the potential singularities in
the tidal forces. Indeed, unlike Lifshitz spacetime, the Lorentzian soliton is
geodesically complete and regular everywhere.

\bigskip

An explicit example of these results was provided by showing the existence of
the gravitational soliton (\ref{LifshitzSoliton}) for BHT massive gravity,
which possesses the precise mass, given by (\ref{M-Soliton}), that
successfully reproduces the entropy of the analytic asymptotically Lifshitz
black hole with $z=3$ found in Ref. \cite{ABGGH00}. In this case, the soliton
is an asymptotically Lifshitz spacetime with $z=$ $\frac{1}{3}$. Note that for
General Relativity, making use of \textquotedblleft reverse
engineering\textquotedblright, i.e., plugging asymptotically Lifshitz metrics
into the Einstein field equations in order to find the corresponding stress
energy tensor, it has been shown that the null energy condition is violated
for $z<1$ \cite{Lifshitz-energy conditions1, Lifshitz-energy conditions2}. It
is then worth to remark that since the the gravitational soliton
(\ref{LifshitzSoliton}) is a solution of the field equations in vacuum, the
whole spacetime is devoid of any kind of stress-energy tensor, and hence no
energy conditions can be violated. Thus, the result found for General
relativity does not apply for BHT massive gravity.

\bigskip

As an ending remark, it would be interesting to explore whether the soliton
mass captures a possible central charge of an affine extension of the Lifshitz
algebra (\ref{algebraLifshitzZ}).

\bigskip

\textit{Acknowledgments.} We thank Pedro Alvarez, G. Barnich, Francisco
Correa, Cristi\'{a}n Mart\'{\i}nez, and specially to Stephane Detournay and
Alfredo P\'{e}rez for many useful and enlightening discussions. This work has
been partially funded by the Fondecyt grants N$%
{{}^\circ}%
$ 1085322, 1095098, 3110141, and by the Conicyt grant ACT-91:
\textquotedblleft Southern Theoretical Physics Laboratory\textquotedblright%
\ (STPLab). H.G. thanks Conicyt for financial support. The Centro de Estudios
Cient\'{\i}ficos (CECs) is funded by the Chilean Government through the
Centers of Excellence Base Financing Program of Conicyt.

\end{document}